\documentclass[aps,prl,twocolumn,preprintnumbers]{revtex4-1}
\usepackage{hyperref}
\usepackage{amsmath}
\usepackage{amsfonts}
\usepackage{bm}
\usepackage{bbold}
\usepackage{xcolor}

\def\nn{\nonumber}

\def\PC{PC}
\def\LS{ND}

\def\eqn#1{eq.~(\ref{#1})}
\def\Eqn#1{Equation~(\ref{#1})}
\def\eqns#1#2{eqs.~(\ref{#1}) and~(\ref{#2})}

\def\braket#1{\langle #1 \rangle}

\def\spa#1.#2{\left\langle#1\,#2\right\rangle}
\def\spb#1.#2{\left[#1\,#2\right]}

\def\be{\begin{equation}}
\def\ee{\end{equation}}
\def\bea{\begin{eqnarray}}
\def\eea{\end{eqnarray}}  
\def\beal{\begin{equation}\begin{aligned}}
\def\eeal{\end{aligned}\end{equation}}
\def\nn{\nonumber}
\def\braket#1{\langle #1 \rangle}

\allowdisplaybreaks

\begin{document} 

\preprint{
NORDITA 2022-091,
UUITP-57/22
}

\title{Kerr Black Holes From Massive Higher-Spin Gauge Symmetry}
\author{Lucile Cangemi$^{1}$}
\email{lucile.cangemi@physics.uu.se}
\author{Marco Chiodaroli$^{1}$}
\email{marco.chiodaroli@physics.uu.se}
\author{Henrik Johansson$^{1,2}$}
\email{henrik.johansson@physics.uu.se}
\author{Alexander Ochirov$^{3,4}$}
\email{ao@lims.ac.uk}
\author{Paolo Pichini$^{1}$}
\email{paolo.pichini@physics.uu.se}
\author{Evgeny Skvortsov$^{5}$}
\email{evgeny.skvortsov@umons.ac.be}
\affiliation{$\mbox{}^{1}$Department of Physics and Astronomy, Uppsala University, Box 516, 75120 Uppsala, Sweden,}
\affiliation{$\mbox{}^{2}$Nordita, Stockholm University and KTH Royal Institute of Technology, Hannes Alfv\'{e}ns  v\"{a}g 12, 10691 Stockholm, Sweden,}
\affiliation{$\mbox{}^{3}$Mathematical Institute, University of Oxford,
Woodstock Rd, Oxford OX2 6GG, UK,}
\affiliation{$\mbox{}^{4}$London Institute for Mathematical Sciences, Royal Institution, 21 Albemarle St, London W1S 4BS, UK,}
\affiliation{$\mbox{}^{5}$Service de Physique de l’Univers, Champs et Gravitation,
Universit\'{e} de Mons, 20 place du Parc, 7000 Mons, Belgium}

\begin{abstract}
We propose that the dynamics of Kerr black holes is strongly constrained by the principle of gauge symmetry. We initiate the construction of EFTs for Kerr black holes of any integer quantum spin $s$ using St{\"u}ckelberg fields, and show that the known three-point Kerr amplitudes are uniquely predicted using massive higher-spin gauge symmetry. This symmetry is argued to be connected to an enhanced range of validity for the Kerr EFTs. We consider the closely related root-Kerr electromagnetic solution in parallel, for which the dynamical interactions with photons are also constrained by massive higher-spin gauge symmetry. Finally, the spin-$s$ Compton amplitudes are analyzed, and we discuss contact-term constraints at $s=2$ from Ward identities.
\end{abstract} 


\maketitle

\section{Introduction}

After the first direct detection of gravitational waves from merging black-hole binaries by the LIGO/Virgo collaboration~\cite{LIGOScientific:2016aoc}, the need for matching theory with experiment has driven the development of novel computational methods~\cite{Damour:2016gwp}. 
Recently, it has become increasingly clear that the problem of classical gravitational radiation is deeply connected with modern quantum approaches. Effective field theory (EFT) posits that well-separated masses admit an effective description in terms of point particles~\cite{Goldberger:2004jt}. While traditional EFT descriptions of black holes employ worldline formalisms~\cite{Goldberger:2004jt,Goldberger:2007hy,Porto:2005ac,Levi:2015msa,Bini:2017xzy,Levi:2018nxp,Mogull:2020sak,Guevara:2020xjx}, more recent approaches directly utilize properties of quantum~\emph{scattering amplitudes}~\cite{Bern:2019nnu,Bern:2021dqo}. Classical limit~\cite{Kosower:2018adc} and conversion to bound systems~\cite{Kalin:2019rwq} allow for relevant information to be extracted, such as the effective two-body potential~\cite{Cheung:2018wkq} (see the review~\cite{Buonanno:2022pgc}).
Spinning Kerr black holes have been studied using amplitude-based methods~\cite{Vaidya:2014kza,Guevara:2017csg,Guevara:2018wpp,Chung:2018kqs,Bautista:2019tdr,Maybee:2019jus,Guevara:2019fsj,Arkani-Hamed:2019ymq,Bern:2020buy,Aoude:2020onz,Chiodaroli:2021eug,Aoude:2021oqj,Chen:2021kxt,FebresCordero:2022jts,Bern:2022kto}; however, the inclusion of higher spin-multipole effects
remains poorly understood~\cite{Bautista:2021wfy,Aoude:2022trd,Bern:2022kto,Aoude:2022thd,Bautista:2022wjf}, due to difficulties in identifying proper higher-spin amplitudes~\cite{Arkani-Hamed:2017jhn}.  

Quantum field theory (QFT) provides a natural framework for exploring scattering amplitudes with spin. Gauge symmetry is essential in covariant formalisms for theories with spin, ensuring both the correct degrees of freedom and mild dependence on scales (e.g.~renormalizability/enhanced Wilsonian cutoff). Evidence from low-spin EFTs~\cite{Johansson:2019dnu,Chiodaroli:2021eug} compatible with known Kerr amplitudes~\cite{Arkani-Hamed:2017jhn,Guevara:2018wpp,Saketh:2022wap} points to spontaneously broken gauge symmetry or enhanced tree-level unitarity properties as useful guiding principles~\cite{Chiodaroli:2021eug}.
We note that spinning black holes may exist with masses ranging all the way down to the Planck scale, suggesting that Kerr EFTs should have enhanced range of validity compared to EFTs of generic compact objects, such as neutron stars.

Finding a suitable tree-level Compton amplitude associated to a Kerr black hole is an important open problem in view of its relevance~\cite{Vaidya:2014kza,Guevara:2017csg} for the two-body problem at ${\cal O}(G_\text{N}^2)$, where $G_\text{N}$ is Newton's constant.
While matching to the Kerr metric fixes the pole part of the Compton amplitude \cite{Guevara:2018wpp,Chung:2018kqs,Arkani-Hamed:2019ymq}, a contact-term ambiguity is present in the opposite-helicity sector~\cite{Arkani-Hamed:2017jhn}. Resolving this ambiguity is an ongoing effort \cite{Falkowski:2020aso,Chung:2018kqs,Chiodaroli:2021eug,Bautista:2021wfy,Bern:2022kto,Aoude:2022trd,Aoude:2022thd,Bautista:2022wjf}, and it may require the identification of new physical principles associated to black-hole scattering. 

In this Letter, we show that known Kerr amplitudes come from EFTs that enjoy massive higher-spin gauge symmetry, and demonstrate that this property is highly constraining when combined with EFT principles, such as low-derivative counting and tree-level unitary considerations. Using St{\"u}ckelberg fields, Ward identities and results from the higher-spin literature~\cite{Zinoviev:2001dt,Zinoviev:2006im,Zinoviev:2009hu}, we investigate the constraints imposed by gauge symmetry at three points, as well as for root-Kerr~\cite{Monteiro:2014cda,Arkani-Hamed:2019ymq} Compton amplitudes. Based on our findings, we conjecture that higher-spin gauge symmetry is a strong selection principle for describing the full dynamics of Kerr black holes. 

At three points, there is now a firmly established relation
\cite{Guevara:2018wpp,Chung:2018kqs,Guevara:2019fsj,Arkani-Hamed:2019ymq,Aoude:2020onz,Aoude:2021oqj}
between the gravitational interaction of a Kerr black hole
and the three-point scattering amplitude
\cite{Arkani-Hamed:2017jhn}
\be \label{KerrAmp}
M_{\rm Kerr}(\Phi_1^s\,\Phi_2^s\,h_3^+)
= M_0 \frac{\braket{\bm{12}}^{2s}}{m^{2s}} ,
\ee
where $\Phi_{1,2}^{s}$ denote spin-$s$ mass-$m$ fields and $h_3^+$ a positive-helicity graviton (parity gives the negative-helicity formula). The spin-0 case
$M_0= -\kappa (p_1\cdot\varepsilon_3^+)^2$ describes a Schwarzschild black hole, where $\kappa=\sqrt{32\pi G_\text{N}}$.

We use on-shell Weyl spinors,
$p_i {\cdot} \sigma |\bm{i}] = m |\bm{i}\rangle$
\cite{Arkani-Hamed:2017jhn}, and consider integer spin-$s$ massive polarization tensors
\cite{Guevara:2018wpp,Chung:2018kqs}
\be
\bm{\varepsilon}^{\mu_1 \mu_2 {\cdots} \mu_s}_i = \bm{\varepsilon}^{\mu_1}_i\bm{\varepsilon}^{\mu_2}_i {\cdots} \bm{\varepsilon}^{\mu_s}_i ,
\qquad \quad \bm{\varepsilon}^{\mu}_i =
\frac{\langle \bm{i}|\sigma^\mu| \bm{i}]}{\sqrt{2}m} ,
\label{PolTensors}
\ee
where $\sigma^\mu$ are the Pauli/van der Waerden matrices. The spin-1 polarization vectors satisfy
$p_i \cdot \bm{\varepsilon}_i= \bm{\varepsilon}_i^2 =0$,
implying transversality and tracelessness of the tensors~\eqref{PolTensors}. 

The Kerr amplitude can be re-expressed in terms of its spin vector $S^\mu = m a^\mu$, where $a^\mu$ is the Kerr ring radius. For quantum spin $s$, we parametrize the spin vector of the incoming black hole (label 1), using complex SU(2) coordinates $z^a$, which can be seen as projective coordinates on the 3-sphere group manifold,
\be
S^\mu = - \frac{s}{2m} (\bar{z}^a z_a)^{2s-1} \big(
  \langle\overline{\bm{1}}| \sigma^\mu |\bm{1}]
+ \langle\bm{1}| \sigma^\mu |\overline{\bm{1}}] \big) .
\ee
Here (and in the above equations), the Weyl spinors are taken to be functions of $z$, namely $|\bm{1}\rangle := |1^a\rangle z_a$,
$|\bm{1}] := |1^a] z_a$,
$|\overline{\bm{1}}\rangle := |1^a\rangle \bar z_a$ and
$|\overline{\bm{1}}] := |1^a] \bar z_a$,
where the un-bold Weyl spinors carry SU(2) little-group indices. By suitable choices of little-group frames, one can relate the incoming and outgoing (label 2) spinors in the three-point amplitudes by a Lorentz transformation,
$| \bm{2} \rangle :=
|\overline{\bm{1}} \rangle + p_3 \cdot\sigma| \overline{\bm{1}}]/(2m)$.

The spin vector is automatically transverse $p_1 \cdot S=0$,
and moreover gives the quantum-mechanical spin operator $\hat S^\mu$ upon acting with derivatives,
\be
(\hat S^\mu)_{\vec a}{}^{\vec b} :=
\frac{1}{(2s)!^2} \Big(\prod_{j=1}^{2s}
\frac{\partial~}{\partial \bar z^{a_j}}
\frac{\partial~}{\partial z_{b_j}}
\Big) S^\mu ,
\ee
with the multi-index notation $\vec a= a_1a_2\cdots a_{2s}$. It follows that $\hat S^2= -s(s+1) \mathbb{1}$ and  $[\hat S^\mu,\hat S^\nu] = i  \epsilon^{\mu \nu\rho} \hat S_\rho$,  with ${\rm SO}(3)$ structure constants $\epsilon^{\mu \nu\rho}= \epsilon^{\mu \nu\rho \sigma} {p_{1\sigma}/m}$. Conversely, the spin vector can be considered as the expectation value of the spin operator $S^\mu = \big\langle \hat S^\mu \big\rangle :=  (\bar  z)^{2s} \cdot \hat S^\mu \cdot  (z)^{2s}$, where the $z^a$ encode the spin quantum state.  

With the introduced variables, the Kerr amplitude \eqn{KerrAmp} is equal to the expectation value of an exponentiated spin operator
\cite{Guevara:2018wpp,Bautista:2019tdr,Guevara:2019fsj}:
\be
M_0 \frac{\braket{\bm{12}}^{2s}}{m^{2s}}
= M_0 \Big\langle \sum_{n=0}^{2s} \frac{1}{n!}
\bigg(\frac{p_3 \cdot \hat S}{m}\bigg)^{\!n} \Big\rangle
= M_0 \big\langle e^{p_3 \cdot \hat a}\big\rangle ,
\ee
where the ring-radius operator is $\hat a^\mu = \hat S^\mu/m$. In the classical limit, $s\rightarrow \infty$, $S^\mu$ finite, the quantum variance vanishes, implying that $\big\langle (\hat a^\mu)^n \big\rangle= \big\langle \hat a^\mu \big\rangle^n= (a^\mu)^n$, and one recovers that the classical amplitude of Kerr is related to Schwarzschild though the Newman-Janis shift~\cite{Newman:1965tw,Arkani-Hamed:2019ymq}
$M_{\rm Kerr} = M_0 e^{p_3 \cdot a}$.

As has been established in several works~\cite{Monteiro:2014cda,Arkani-Hamed:2019ymq,Guevara:2020xjx}, there is a closely related electromagnetic root-Kerr solution, describing a massive spinning state $\Phi^s$ with charge $Q$ coupled to a photon $A^\mu$. The corresponding three-point amplitude~\cite{Arkani-Hamed:2017jhn} is very similar to the one for Kerr:
\be \label{RootKerrAmp}
A_{\!\sqrt{\rm Kerr}}(\Phi_1^s\, \overline{\Phi}_2^s\, A_3^+)
= A_0 \frac{\braket{\bm{12}}^{2s}}{m^{2s}}
= A_0 \big\langle e^{p_3 \cdot \hat a} \big\rangle , 
\ee
where $A_0 = 2Q\,p_1\cdot\varepsilon_3^+$ is a charged-scalar amplitude.

\section{Kerr and root-Kerr EFT${\rm s}$}

The infinite family of three-point Kerr amplitudes gives reasons to expect the existence of an underlying QFT framework. One may anticipate two families of Kerr and root-Kerr EFTs parametrized by the quantum spin $s$. Supporting evidence is provided by the existence of well-behaved low-spin quantum Compton amplitudes (up to spin-2 Kerr and spin-1 root-Kerr~\cite{Arkani-Hamed:2017jhn}) that agree with classical Kerr results~\cite{Saketh:2022wap}. In ref.~\cite{Chiodaroli:2021eug}, it was also argued that the corresponding low-spin EFTs are determined using tree-level unitarity constraints, spontaneous symmetry breaking, and restrictive derivative counting. 

In the gravitational Kerr case, the EFTs for $s=1/2,1$ and $3/2$ are minimally coupled~\cite{Arkani-Hamed:2017jhn} Majorana spinor, Proca and Rarita–Schwinger fields, respectively. For the spin-2 Kerr case, the interactions are compatible with a Kaluza-Klein graviton~\cite{Chiodaroli:2021eug}. For the gauge-theory root-Kerr case, $s=1/2$ is a minimally coupled Dirac fermion, and $s=1$ is a $W$-boson. Furthermore, for spin-3/2 root-Kerr and spin-5/2 Kerr, these theories admit a higher-spin current that is conserved up to mass terms, $\partial \cdot  J = {\cal O}(m)$ \cite{Ferrara:1992yc,Porrati:1993in,Cucchieri:1994tx,Chiodaroli:2021eug}, which fixes their Lagrangians and provides a conjecture for the corresponding Compton amplitudes. The spin-3/2 case is a non-minimally coupled gravitino \cite{Deser:2000dz,Zinoviev:2009hu} that can be embedded in a theory that exhibits spontaneously broken supersymmetry (i.e. a gauged supergravity~\cite{Freedman:1976aw}), and the spin-5/2 theory has also been featured in the literature~\cite{Porrati:1993in,Cucchieri:1994tx,Chiodaroli:2021eug}. 
The low-spin Kerr and root-Kerr theories are summarized in Table~\ref{tabKerr}.
We will now demonstrate how to fix the cubic interactions in theories with arbitrary integer spin, and moreover constrain the contact terms in the Compton amplitude for the spin-2 root-Kerr case.

\begin{table}
\begin{tabular}{|c||c|c|c|c|c|c|} 
 \hline
 EFTs & $s=\,^1\!/\!_2$ & $s=1$ & $s=\,^3\!/\!_2$ & $s=2$ & $s=\,^5\!/\!_2$ & $s \ge 3$ \\
 \hline
 \hline
Kerr & Major. & Proca & Rar.-Sch. & KK\,grav. & \cite{Chiodaroli:2021eug} & HS \\ [0.5ex]  \hline
${\sqrt{\rm Kerr}}$ & Dirac & $W$\!-boson & gravitino & HS & - & HS \\ [0.5ex] 
 \hline
\end{tabular}
\caption{\small \it Kerr and $\sqrt{\text{Kerr}}$ theories; the unknown higher-spin (HS) theories for integer $s$ are the subject of this Letter. \label{tabKerr}}
\vskip -10pt
\end{table}

\smallskip
\noindent
{\bf Spin-2 root-Kerr theory:} Let the complex fields $\{\Phi_{\mu\nu},B_\mu,\varphi\}$ collectively describe the charged root-Kerr matter, of which $\Phi_{\mu\nu}$ is the physical (symmetric) spin-2 field and $\{B_\mu,\varphi\}$ are St\"uckelberg auxiliary fields.
The linearized gauge transformations of the massive fields are~\cite{Zinoviev:2001dt,Zinoviev:2006im,Zinoviev:2009hu} 
\beal \label{spin2gaugetr}
\delta \Phi_{\mu\nu} & = \frac{1}{2}\partial_\mu \xi_\nu + \frac{1}{2}\partial_\nu \xi_\mu + \frac{m}{\sqrt{2}} \eta_{\mu\nu} \zeta, \\ \delta B_\mu & = \partial_\mu \zeta + \frac{m}{\sqrt{2}} \xi_\mu , \\ \delta \varphi & = \sqrt{3} m \zeta ,
\eeal
where $\{\xi_\mu, \zeta\}$ are (complex) gauge parameters, the metric is $\eta_{\mu\nu}={\rm diag}(+,-,-,-)$, and field normalizations are conveniently adjusted~\cite{Zinoviev:2001dt} to match our spin-$s$ discussion. It is straightforward to find a unique free Lagrangian ${\cal L}_0$ satisfying $\delta{\cal L}_0=0$, but let us postpone its introduction.

Instead, we consider three-point Ward identities when the root-Kerr matter is coupled to a (massless) photon $A^\mu$. We work in momentum space, and represent all fields as off-shell plane waves with momenta $p_i$ and unconstrained  polarizations $\Phi^{\mu\nu}_i{\sim}\,\epsilon^\mu_i \epsilon^\nu_i$,
$B^\mu_i{\sim}\,\epsilon^\mu_i$,
$\xi^\mu_i{\sim}\,\epsilon^\mu_i$, $ A^\mu_i{\sim}\,\epsilon^\mu_i$,
only distinguished by their particle-label subscripts. Let $V_{\Phi\overline\Phi A}(\epsilon_i, p_i)$, $V_{B \overline \Phi A}(\epsilon_i, p_i)$, etc., denote off-shell vertex functions, then the linearized gauge transformation~\eqn{spin2gaugetr} translates into
\begin{align} \label{eq10}
V_{\xi \overline \Phi A} & := \frac{m}{\sqrt{2}} V_{B \overline \Phi A} -\frac{i}{2}p_1 {\cdot} \frac{\partial~}{\partial \epsilon_1} V_{\Phi  \overline\Phi A} , \\
V_{\zeta \overline \Phi A} & := \sqrt{3}m V_{\varphi \overline \Phi A} -i p_1   {\cdot}  \frac{\partial~}{\partial \epsilon_1} V_{B \overline \Phi A}  + \frac{m}{2\sqrt{2}}   \bigg(\!\frac{\partial~}{\partial \epsilon_1}\!\bigg)^{\!2} V_{\Phi \overline \Phi A} . \nn
\end{align}
and the Ward identities are $V_{\xi \overline \Phi A}\big|_{(2,3)}{=}V_{\zeta \overline \Phi A}\big|_{(2,3)}{=}0$ where the restriction means that they are evaluated on shell for legs 2 and 3, $p_2^2-m^2=p_3^2=p_i\cdot \epsilon_i =\epsilon_i^2=0$.

To find the interactions that satisfy the Ward identities, we write down Ans\"{a}tze for the vertex functions.
We find that sufficiently large Ans\"{a}tze have the following schematic derivative counting:
\beal
V_{\Phi \overline \Phi A} & \sim m\, (\epsilon_1)^2 \, (\epsilon_2)^2 \, \epsilon_3\, \Big(\frac{p^3}{m^3}+\frac{p}{m}\Big), \\
V_{B \overline \Phi A} & \sim m\, (\epsilon_1) \, (\epsilon_2)^2 \, \epsilon_3\, \Big(\frac{p^2}{m^2}+1\Big), \\
V_{\varphi \overline \Phi A} & \sim m\,  (\epsilon_2)^2 \, \epsilon_3\, \Big(\frac{p}{m}\Big),
\eeal
which in hindsight is compatible with interactions with $\partial^{s_1+s_2-1}$ derivatives, between massive spin-$s_1$ and spin-$s_2$ fields and a spin-1 massless boson. The terms linear in momentum typically come from the minimally coupled theory ${\cal L}_0$, obtained by covariantizing the free theory $\partial_\mu\rightarrow D_\mu =\partial_\mu + i Q A_\mu$, and the higher-derivative terms must come from non-minimal couplings involving the abelian field strength $F_{\mu\nu}=\partial_{\mu}A_\nu-\partial_{\nu}A_\mu$. Thus we constrain the Ans\"{a}tze such that all terms non-linear in momenta involve $p_3^{[\mu}\epsilon_3^{\nu]}$. 

Imposing the Ward identities on the vertex functions and plugging in on-shell variables $\epsilon_i^\mu\rightarrow {\bm \varepsilon}_i^\mu$ gives the following amplitude
\begin{equation}
A(\Phi_1^{s{=}2}\, \overline{\Phi}{}_2^{s{=}2} A_3^+)\, {=}\, 
 A_0 \frac{\braket{\bm{12}}^3}{m^4} \!\big(c_1  [\bm{12}] + (1 {-} c_1 ) \braket{\bm{12}}\big),
 \end{equation}
where $c_1$ is a free parameter. A unique amplitude is obtained after imposing a current constraint (tree-level unitarity constraint \cite{Chiodaroli:2021eug})
\begin{align}\label{eq:currentconstraint}
& p_1 {\cdot} \frac{\partial~}{\partial \epsilon_1}
V_{\Phi \overline \Phi A} \Big|_{(2,3)}={\cal O}(m).
\end{align}
This gives $c_1=0$ and thus precisely the root-Kerr amplitude~\eqref{RootKerrAmp}.

\noindent
{\bf Spin-3~Kerr~theory:} Let the real, symmetric-tensor, fields $\{\Phi_{\lambda\mu\nu},H_{\mu\nu},B_\mu,\varphi\}$
collectively describe the Kerr black hole, where $\Phi_{\lambda\mu\nu}$ is the physical spin-3 field and $\{H_{\mu\nu},B_\mu,\varphi\}$ are St\"{u}ckelberg auxiliary fields.
The linearized gauge transformations are~\cite{Zinoviev:2001dt, Zinoviev:2006im,Zinoviev:2009hu}
\beal
\delta \Phi_{\lambda\mu\nu} & =  \partial_{(\lambda} \xi_{\mu\nu)} + \frac{\sqrt{3}}{4} m \eta_{(\lambda\mu} \xi_{\nu)} , \\ \delta H_{\mu\nu} & = \partial_{(\mu} \xi_{\nu)} + \frac{m}{\sqrt{3}} \xi_{\mu\nu}
+  \frac{\sqrt{5}}{2}  m \eta_{\mu\nu} \xi, \\ \delta B_\mu & = \partial_\mu \xi + \frac{\sqrt{5}}{2}m \xi_\mu , \\ \delta \varphi & = \sqrt{6} m \xi ,
\eeal
where $\xi_{\mu\nu}$ is a traceless symmetric tensor. A unique free Lagrangian ${\cal L}_0$ invariant under the gauge transformation can be constructed, but let us first consider three-point Ward identities. We write up Ans\"{a}tze for the off-shell vertices $V_{\Phi\Phi h}$, $V_{H\Phi h}$, $V_{B\Phi h}$, $V_{\varphi\Phi h}$, where $h$ corresponds to the graviton field $h^{\mu\nu}$. Similar to our previous analysis, we find that sufficiently large Ans\"{a}tze have at most $\partial^{s_1+s_2-2}$ derivatives, where $s_2=3$ and $s_1$ is the spin of the first field in each vertex. The terms linear and quadratic in derivatives are chosen so they are compatible with the covariantization of the free Lagrangian $\sqrt{-g}{\cal L}_0$, where $\partial_\mu \rightarrow \nabla_\mu$ and $\eta_{\mu \nu} \rightarrow g_{\mu \nu} =\eta_{\mu \nu}+  \kappa h_{\mu \nu}$,  which can be checked through the massless Ward identities of the graviton. For the higher-derivative terms we demand that they are built out of a linearized Riemann tensor, which on shell becomes $R^{\mu\nu\rho\sigma}\sim F^{\mu\nu}F^{\sigma\rho}\sim p_3^{[\mu}\epsilon_3^{\nu]} p_3^{[\rho}\epsilon_3^{\sigma]}$, where the polarization of the graviton is $h^{\mu \nu}=\epsilon_3^{\mu} \epsilon_3^{\nu}$.   

Writing all the vertices as functions of $\{\epsilon^\mu_i, p_i^\mu\}$, the Ward identities become three functional constraints,
\begin{align}
& \frac{m}{\sqrt{3}} V_{H \Phi h}
- \frac{i}{3}p_1 {\cdot} \frac{\partial~}{\partial \epsilon_1} V_{\Phi \Phi h}\bigg|_{(2,3),\epsilon_1^2\rightarrow 0}\!\!= 0 ,  \\
& \frac{\sqrt{5}}{2}m V_{B \Phi h}
  - \frac{i}{2}p_1 {\cdot} \frac{\partial~}{\partial \epsilon_1} V_{H \Phi h}
+ \frac{m}{8\sqrt{3}} \bigg(\!\frac{\partial~}{\partial \epsilon_1}\!\bigg)^{\!2}
  V_{\Phi \Phi h}
  \bigg|_{(2,3)}\!\!\!\!\!= 0 , \nn \\
& \sqrt{6}m V_{\varphi \Phi h}
- ip_1 {\cdot} \frac{\partial~}{\partial \epsilon_1} V_{B \Phi h}
+ \frac{\sqrt{5}}{4}m
  \bigg(\!\frac{\partial~}{\partial \epsilon_1}\!\bigg)^{\!2} V_{H \Phi h}\bigg|_{(2,3)}\!\!\!\!\!= 0 , \nn
\end{align}
where as before the restriction implies that legs $(2,3)$ are placed on shell. Additionally, we set $\epsilon_1^2\rightarrow 0$  given that $\xi^{\mu\nu}$ is traceless. 

After imposing the Ward identities on the Ans\"{a}tze, and imposing the on-shell conditions, we obtain a unique amplitude exactly matching Kerr,
\be 
M(\Phi_1^{s{=}3}\,\Phi_2^{s{=}3}\,h_3^+)
= M_0 \frac{\braket{\bm{12}}^{6}}{m^{6}} \,.
\ee
In this case, the current constraint \eqref{eq:currentconstraint} is not required for uniqueness, yet it is consistent with the solution.

\noindent
{\bf Spin-$s$ Kerr and root-Kerr EFTs:}
For the spin-$s$ massive theories we follow Zinoviev's~\cite{Zinoviev:2001dt} approach of introducing a tower $k=0,1,2,\ldots,s$ of tensor fields and gauge parameters, 
\be
\Phi^k := \Phi^{\mu_1 \mu_2 \cdots \mu_k} , ~~~~~~~ \xi^k := \xi^{\mu_1 \mu_2 \cdots \mu_k} ,
\ee
where all fields are symmetric tensors, and the gauge parameters are traceless ${\rm Tr\,}\xi^k=0$, and the fields are double-traceless ${\rm Tr\,}\tilde \Phi^k=0$, where the trace is denoted by $\tilde \Phi^k:={\rm Tr\,}\Phi^k$. In terms of these fields, the linearized gauge transformation can be compactly written as
\be \label{linearisedDeltaPhi}
\delta \Phi^k = \partial^{(1} \xi^{k-1)} + m \alpha_k \xi^k + m \beta_k \eta^{(2} \xi^{k-2)} ,
\ee
where the $k$ distinct Lorentz indices of the fields, derivative $\partial^1:=\partial^{\mu_1}$, and metric $\eta^2:=\eta^{\mu_1 \mu_2}$ are to be symmetrized, as indicated. The numbers $\alpha_k$ and $\beta_k$ are~\cite{Zinoviev:2001dt}
\begin{equation}
\alpha_k = \frac{1}{k {+} 1}\sqrt{\frac{(s {-} k) (s {+} k {+} 1)}{2}}, \quad
\beta_k = \frac{1}{2}\frac{k}{k {-} 1} \alpha_{k - 1} ,
\end{equation}
such 
that the highest-spin gauge parameter $\xi^s$ decouples and thus $\Phi^s$ carries the physical degrees of freedom. 

The free theories $\mathcal{L}_{0}$ that are invariant under the gauge transformation~\eqref{linearisedDeltaPhi} can be decomposed into a Feynman-gauge part ${\cal L}_{\rm F}$ and a sum over gauge-fixing functions $G^k$,
\begin{equation} \label{FreeL0}
 \mathcal{L}_0 = {\cal L}_{\rm F} + \frac{1}{2}\sum_{k=0}^{s-1} (-1)^{k} (k+1) G^k G^k .
\end{equation}
The Feynman-gauge part is completely diagonal 
\begin{equation}
  {\cal L}_{\rm F} \!= \! -\!\!\sum_{k=0}^s\!
  \frac{({-}1)^k}{2}\!\bigg[ \Phi^k (\Box + m^2) \Phi^k - \frac{k(k{-}1)}{4} \tilde{\Phi}^k(\Box + m^2)\tilde{\Phi}^k \! \bigg],
\end{equation} 
and the $G^k$s give rise to off-diagonal contributions
\be
G^k\! \!=\!\partial {\cdot} \Phi^{k+1}  {-} \frac{k}{2} \partial^{(1}  \tilde{\Phi}^{k{+}1)} {+}  m \big(\alpha_{k} \Phi^k {-}   \gamma_{k}  \tilde{\Phi}^{k{+}2} 
{-}  \delta_k \eta^{(2}  \tilde{\Phi}^{k)}\big),
\ee
where $\gamma_k = \frac{1}{2}(k+1) \alpha_{k+1}$ and $\delta_{k}= \frac{1}{4}(k-1) \alpha_{k}$.

\def\h{\mathfrak{h}}
In order to introduce interactions, we proceed as before and write down Ans\"{a}tze for the  relevant off-shell vertices
\be
V_{\Phi^{k} \Phi^s A^\h} = V^\text{min.}_{\Phi^{k} \Phi^s A^\h} +V^\text{non-min.}_{\Phi^{k} \Phi^s A^\h} ,
\ee
where the minimal interactions are uniquely determined by the covariantization of ${\cal L}_0$, and the non-minimal interactions are built out of the field strength $(F_{\mu \nu})^\h$, and here we treat Kerr $(h^{\mu \nu}{:=}A^{\h=2})$ and root-Kerr $(A^\mu{:=}A^{\h=1})$ simultaneously using the helicity parameter $\h=1,2$. We will suppress the complex nature of the root-Kerr fields, but it is understood that they carry U(1) charge $Q$.

Using the linearized gauge transformation \eqref{linearisedDeltaPhi} one can define  gauge-transformed off-shell vertices
\beal
V_{\xi^k \Phi^s  A^\h} := m \alpha_k  V_{\Phi^k \Phi^s A^\h}
- \frac{i}{k+1}   p_1 {\cdot} \frac{\partial~}{\partial \epsilon_1} V_{\Phi^{k+1} \Phi^s A^\h} &\!\!\\
+\,\frac{m \beta_{k+2}}{(k+2)(k+1)}
  \frac{\partial~}{\partial \epsilon_1} {\cdot} \frac{\partial~}{\partial \epsilon_1}
  V_{\Phi^{k+2} \Phi^s A^\h} &.\!\!\!
\eeal
The constraints that we impose on the interactions can then be summarized as:
\begin{itemize}
\item[(MC)] Minimal-coupling extension of ${\cal L}_0$ gives $V^\text{min.}_{\Phi^{k} \Phi^s A^\h}$;
\item[\rlap{(WI)}\hphantom{(MC)}]
Ward identities~$V_{\xi^k \Phi^s  A^\h} \big|_{(2,3),\epsilon_1^2\rightarrow 0}=0$;
\item[\rlap{(CC)}\hphantom{(MC)}] Current~constraint $p_1{\cdot} \frac{\partial~}{\partial \epsilon_1}V_{\Phi^{s} \Phi^{s} \!  A^\h}\! \big|_{\! (2,3),\epsilon_1^2\rightarrow 0}\!\! = \! {\cal O}(m)$.
\end{itemize}
Additionally, considering the lowest-derivative solutions that satisfy the above constraints give refined Ans\"{a}tze:  
\begin{itemize}
\item[\rlap{(PC)}\hphantom{(MC)}] Power-counting bound on derivatives in non-minimal vertices: $V^\text{non-min.}_{\Phi^{s_1} \Phi^{s_2} A^\h}~\sim~\partial^{s_1+s_2-2\h}(F_{\mu\nu})^\h$;
\item[\rlap{(\LS)}\hphantom{(MC)}] Near-diagonal interactions: if $|s_1{-}s_2|>\!\h$ then $V_{\Phi^{s_1}\Phi^{s_2}A^{\h}} =0$.
\end{itemize}

\def\spa12{\langle 12 \rangle}
\def\spb12{[12]}

Considering spin-$s$ root-Kerr theory with conditions (MC)+(\PC)+(WI), we obtain the three-point amplitude
\begin{equation} \label{Unfixed_Root_Kerr}
A(\Phi_1^s\,\overline\Phi{}_2^s\,A_3^+)  = 
 A_0 \frac{ \langle \bm{12}\rangle^{2s}}{m^{2s}}
 \left\{ 1 +  \sum_{k=1}^{s-1} c_k  \left( \frac{[\bm{12}]^k}{\langle \bm{12}\rangle^k}
  - 1\right) \right\} ,
\end{equation}
where $c_k$ are unconstrained parameters from the non-minimal Ans\"{a}tze.
Imposing (\LS) constrains the off-shell vertex, but not \eqn{Unfixed_Root_Kerr}, whereas (CC) generates one constraint, $\sum_k c_k=0$. However, imposing both (CC)+(\LS) fixes $c_k=0$, thus uniquely predicting the amplitude~\eqref{RootKerrAmp}. 

Considering spin-$s$ Kerr theory with conditions (MC)+(\PC)+(WI), we obtain the three-point amplitude
\begin{equation} \label{Unfixed_Kerr}
M(\Phi_1^s\, \Phi_2^s\, h_3^+)\!= \!M_0 \frac{ \langle \bm{12}\rangle^{2s}}{m^{2s}}\!\!
 \left\{ 1 {+}  \left(\!1 {-}\frac{[\bm{12}]}{\langle \bm{12}\rangle} \right)^{\!\!2} \sum_{k=0}^{s-4}
   c_k'  \frac{[\bm{12}]^k}{\langle \bm{12}\rangle^k}\! \right\}\!,
\end{equation}
where $c_k'$ are unconstrained parameters from the non-minimal Ans\"{a}tze.
As in the gauge theory, further imposing (\LS) leaves \eqn{Unfixed_Kerr} unchanged.
However, in this case imposing either (CC) or (CC)+(\LS) generates the unique solution $c_k'=0$, thus uniquely predicting the Kerr amplitude~\eqref{KerrAmp}. The above calculations, \eqns{Unfixed_Root_Kerr}{Unfixed_Kerr}, were explicitly carried out through $s\le6$ in root-Kerr and Kerr theories, and beyond this the results are conjectural --- based on the robust patterns we observed. 

We note that the Ward identities can be extended to non-linear gauge invariance for the off-shell three-point functions, which requires Ans\"{a}tze for the non-linear parts of the variation $\delta {\cal L}=0$. We have implemented this up to $s\le4$, and we find that the (WI)+(CC)+(ND) constraints (on the free coefficients in the amplitudes) are now superseded by the stronger requirement of non-linear gauge invariance. More details will be given in ref.~\cite{Cangemi:2023ysz}.

\section{From quantum to classical Compton}

Perturbation theory beyond three points requires that we work out propagators in some simple gauge. The Feynman-gauge propagator $\Delta^{(s)}$ for a spin-$s$ field (physical or auxiliary) is diagonal and has trivial momentum dependence due to the simplicity of ${\cal L}_{\rm F}$, and it acts as a double-traceless projector. It is unique and we find that it is given by the generating function  
\begin{equation}\label{full_prop}
\Delta(\epsilon,\bar \epsilon)=\!\sum_{s=0}^{\infty} (\epsilon)^s {\cdot} \Delta^{\!(s)} {\cdot}  (\bar\epsilon)^s=
      \frac{1}{p^2{-}m^2{+}i0}  \frac{1 - \frac{1}{4}\epsilon^2 \bar \epsilon^2 }{1 + \epsilon \cdot \bar \epsilon + \frac{1}{4} \epsilon^2 \bar \epsilon^2 },
\end{equation}
where, as before, $\epsilon_\mu,\bar \epsilon_\mu$ are auxiliary vectors. 

\def\T{U}
\def\S{V}

Given the root-Kerr EFTs we described, we can now compute the opposite-helicity Compton amplitude (up to contact terms), and we obtain the manifestly local form
\begin{align} \label{Compton}\!\!\!\!\!
A(\Phi_1^s \Phi_2^s A_3^-\!A_4^+) = &\,
\frac{\langle 3|1|4]^2(\T {+} \S)^{2s}\!}{m^{4s}t_{13} t_{14}}
+ \frac{ \langle 3|1|4] \braket{{\bf1}3}[{\bf2}4] P^{(2s)}\!}{m^{4s}t_{13}}~\nn \\ & \null
+ \braket{{\bf1}3} \braket{3{\bf2}} [{\bf1}4] [4{\bf2}] \frac{P^{(2s-1)}\!}{m^{4s}} + C_s , 
\end{align} 
where $t_{ij}=2p_i\cdot p_j$ and $P^{(k)}=\frac{1}{2\S}\big\{(\T + \S)^{k}-(\T - \S)^{k}\big\}$ is a degree-$(k{-}1)$ polynomial in the two local variables $\S = \frac{1}{2}\big(\langle{\bf 1}|4|{\bf 2}] + \langle{\bf 2}|4|{\bf 1}]\big)$,
$\T = \frac{1}{2}\big(\langle{\bf 1}|4|{\bf 2}] {-} \langle{\bf 2}|4|{\bf 1}]\big)- m [{\bf 12}]$. 
The amplitude contains undetermined contact terms $C_s$, where $C_{s\le 3/2}=0$ agrees with the local Compton amplitudes of refs.~\cite{Arkani-Hamed:2017jhn,Chiodaroli:2021eug}.
\Eqn{Compton} provides a novel compact spurious-pole free result that is useful for exposing the remaining contact-term freedom \cite{Arkani-Hamed:2017jhn,Chung:2018kqs,Falkowski:2020aso}. 
For $s=2$ we used Ward identities to narrow down the unfixed $C_2$ to three terms; one is a quantum contribution, and the remaining two are:
\be
C_2\!=\!\frac{\braket{{\bf1}3} \braket{3{\bf2}} [{\bf1}4] [4{\bf2}]}{m^{6}}   \Big\{\!c_1(\braket{{\bf12}}+[{\bf12}])^2+c_2 (\braket{{\bf12}}-[{\bf12}])^2\!\Big\},
\label{ComptonContact}
\ee
where the coefficients $c_i$ are to be determined.
We checked that the same spin-2 amplitude~\eqref{Compton}--\eqref{ComptonContact} can be obtained using the approach of a chiral massive higher-spin Lagrangian~\cite{Ochirov:2022nqz}; the details will be given in ref.~\cite{Cangemi:2023ysz}.

As before, we can re-express the Compton amplitude~\eqref{Compton} in terms of the ring-radius spin operator $\hat a^\mu$, giving the operator-valued amplitude
\begin{align} \label{OperatorAmpl}
-e^{\hat a\cdot q_\perp}\!\bigg(\! \frac{(p_1\cdot \chi)^2}{(p_1 \cdot q_\perp)^2} & - \frac{(p_1\cdot \chi) (\hat a \cdot \chi)}{(p_1 \cdot q_\perp)} + \frac{1}{2s} (\hat a  \cdot \chi)^2\!\bigg) \nn \\*
~~ \null +\,\hat C_s  + {\cal O}(\hat a^2) & + {\cal O}(\hbar) ,
\end{align}
where $q_{\perp} = p_4{-}p_3$, $\chi^\mu = \langle 3|\sigma^\mu|4]$, and we use the small-$\hbar$ scalings $\chi, p_3,p_4 \sim \hbar$ and $  z^a,\bar z^a \sim \hbar^{-1/2}$ (with constraint $\bar{z}^a z_a=1$) to distinguish between the displayed expressions and the  terms ${\cal O}(\hat a^2) + {\cal O}(\hbar)$. A proper classical limit should, however, involve $s\rightarrow \infty$
(or coherent-state sum over all $s$ \cite{Aoude:2021oqj}), in which case the $\hat a^2\sim \hbar^2 s(s+1)$ terms are important. We note that the third term of \eqn{OperatorAmpl} depends on $s$ and thus exhibits spin non-universality starting with the spin quadrupole (see also refs.~\cite{Aoude:2022trd,Cangemi:2022abk}). However, the $c_1$-dependent term in $\hat C_2$ also contains a spin quadrupole that contributes in the $s\rightarrow \infty$ limit, whereas the $c_2$-dependent term starts at the hexadecapole level. More details will be given in ref.~\cite{Cangemi:2023ysz}.

\section{Conclusions}

In this Letter, we propose that the interactions of spinning black holes are strongly constrained by massive higher-spin gauge symmetry involving St{\"u}ckelberg fields. We have shown that the known Kerr and root-Kerr three-point amplitudes come from higher-spin EFTs obeying: Ward identities, low-derivative counting, and a current constraint (used in ref.~\cite{Chiodaroli:2021eug}). 
Starting from a free spin-$s$ formalism~\cite{Zinoviev:2001dt},
we have derived novel spin-$s$ Feynman-gauge Lagrangian and propagators, and considered the minimal and non-minimal interactions. While the non-minimal interactions are strongly constrained by the massive gauge symmetry, more work is needed for obtaining presentable forms. We used the introduced framework to study opposite-helicity Compton amplitudes in the root-Kerr theory, giving a new spin-$s$ formula. Undetermined contact terms start at $s=2$, and we partially constrain them using Ward identities. Initial checks of non-linear gauge invariance and higher-order interactions suggest that gauge invariance and QFT methods are well suited for describing more general dynamics of Kerr black holes, which may include absorption and decay effects.

\section*{Acknowledgements}
We thank F.~Alessio, M.~Ben-Shahar, Z.~Bern, A.~Edison, K.~Haddad, D.~O'Connell, R.~Roiban, O.~Schlotterer, F.~Teng, J.~Vines and Y.~Zinoviev for enlightening discussions related to this work. This research is supported in part by the Knut and Alice Wallenberg Foundation under grants KAW 2018.0116 (From Scattering Amplitudes to Gravitational Waves) and KAW 2018.0162, the Swedish Research Council under grant 621-2014-5722, and the Ragnar S\"{o}derberg Foundation (Swedish Foundations’ Starting Grant). The work of M.C. is also supported by the Swedish Research Council under grant 2019-05283.
The research of A.O. while at Oxford was funded by the STFC grant ST/T000864/1. The work of E.S. was partially supported by the European Research Council (ERC) under the European Union’s Horizon 2020 research and innovation programme (grant agreement No. 101002551) and by the Fonds de la Recherche Scientifique -- FNRS under Grant No. F.4544.21.

\bibliographystyle{apsrev4-1}
\bibliography{references}
\end{document}